# Are Financial Markets an aspect of Quantum World?

Ovidiu Racorean

e-mail: decontatorul@hotmail.com

**Abstract**

Writing the article "Time independent pricing of options in range bound markets"*, the question in the title came naturally to my mind. It is stated, in the above article, that in certain market conditions the stock price is subjected to an equation that exactly matches a time independent Schrodinger equation. The time independent equation for options valuation is used further to explain a stock market phenomenon that resembles an α particle decay tunneling effect. The transmission coefficient for the stock price tunneling effect it is also deduced. Although, it may not have important impact in quantum physics, the philosophical aspects residing in the use of quantum mechanics for stock market specific are very important.

*see the article at: http://arxiv.org/abs/1304.6846

## 1. Introduction

What if the quantum effects could be seen with our very own eyes? It seems to be hardly possible if not totally impossible. The article "Time independent pricing of options in range bound markets" looks to argue with the common sense of understanding quantum mechanics. The article is pure financial but the findings are concerning quantum physics at the philosophical level and at a second look it could be regarded as a new application field for time independent Schrodinger equation.

Starting with the beginning, it should be said that famous Black-Scholes formula for options valuation is the foundation of the article. Separation of variables for pricing options formula lead to deduce a time independent equation for stock prices being in range bound. Specification of range bound market is important since prices moves for a period of time only between support and resistance levels and replicates the behavior of a particle in a box. With this specification in mind, should not be a surprise that the deduced time independent equation for stock prices exactly matches the time independent Schrodinger equation. The differences in the form of the equation come from the specific of stock market which uses notions of volatility and interested rate in place of masses and energies.

Support and resistance levels in market are the walls for the stock price. Stock price "explosive" penetration out of these walls is a financial phenomenon very often seen in the market. With the use of the time independent equation this particular price move is explained as a tunneling effect. It may seem odd to talk about tunneling effect in stock market but it looks reasonably for the mathematical point of view. More, because the shape of the potential in the time independent equation is approximately the same as in nuclear σ particle tunneling, a transmission coefficient relation may be considered.

Transmission coefficient is deduced in the fashion of Gamow α particle decay theory. The relation for T contains also observables from financial markets such as volatility and interest rate, but still can be viewed as a probability, this time, for stock price to tunnel out of the range bound. Discussions of how high the probability should be to allow price to penetrate out, are also related to the width of the wall price should penetrate through. The lowest the walls width will be, the higher the transmission coefficient. The width of walls is low if the stock volatility is low. The price tunneling effect is related with a dramatic fall in the stock volatility right prior a penetration happened. This effect is specific for finance and has not a counterpart in nuclear physics.

Philosophical aspects for the science epistemology are important since it raise numerous legitimate questions for quantum aspects of life. Does quantum effects are not anymore the domain of atomic world? Are financial markets just another aspect of Nature, and subject of the laws of physical world? Or, how "economic" the Nature is, and applies the same lows to similar phenomena, even coming from totally different scientific domains? These are basic questions that at some points are expecting pertinent answers capable to put forward the human kind knowledge.

Although, this article review have not the intention to ask any of this epistemological questions it represents an interesting exercise to bring in the science attention aspects of Nature that deserve a closer look, even the common sense may speak of the impossibility of such an odd approach to a phenomenology until yesterday specific to quantum world.

## 2. The time independent pricing of options equation

Pricing of options commonly is an exercise applying Black-Scholes equation to stock market (and not only) price moves. What happens if the market is in range bound and price move only between support and resistance levels? This price move resembles the 1-D particle in a box case in quantum physics, governed by a time-independent Schrodinger equation. Can such a time independent equation be fined for stock market?

One of the many manners the time-independent Schrodinger equation is deduced, employ a separation of variables in the time dependent form of the equation. The two resulting equations are equated with the same constant E. The spatial dependent equation is the time independent Schrodinger equation. The remaining time dependent equation is used to deduce the constant E, which is found to be the energy.

In the article "Time independent pricing of options in range bound markets", this reasoning is used, having the Black-Scholes formula at the base, in the idea of deduce a time independent equation for prices of stocks that are in range bound. By separation of variables Black-Scholes equation reduces to a simple system of two equations both being equal to the same constant, noted **λ**. Notes in separation of variables for Black-Scholes formula can be found in [1].

As it can be simply deduced, **λ** constant is an important ingredient in the mathematical logic of the article. As in the case of the Schrodinger equation, **λ** constant is deduced from the time dependent equation, which represents the evolution of the option value along with time. Market and macro-economic logic leads to the conclusion that **λ** constant must be of the form:

$$\lambda = \frac{r}{\sigma} \tag{1}$$

where, r is the interest rate and σ stands for the stock volatility.

I am not enter here in much details regarding options valuation and stock market observables since is not the subject of this review; the interested reader can found valuable information in [2] and [3], and other internet resources.

What I will remark here is the price dependent equation that with the **λ** constant value just deduced became:

$$-\frac{\sigma^4}{r(\sigma^2+r)} \frac{d^2\psi_{(S)}}{dS^2} + \frac{1}{S^2}\psi_{(S)} = \frac{r}{\sigma}\psi_{(S)} \tag{2}$$

where S represent the stock price and $\psi_{(S)}$ the function of option valuation. Going further and noting $V_{(S)} = \frac{1}{S^2}$ as the potential equation (2) can be written now as:

$$-\frac{\sigma^4}{r(\sigma^2+r)} \frac{d^2\psi_{(S)}}{dS^2} + V_{(S)}\psi_{(S)} = \frac{r}{\sigma}\psi_{(S)} \tag{3}$$

The resemblance with time-independent Schrodinger equation is obvious.

Interesting to be noted is the shape of the potential $V_{(S)}$ that can be seen in Figure 1 below, along with different levels of **λ** constant.

Figure 1.

Figure 1 is more or less, taking into account the financial market particularities, an exact replica of particle energy in a nuclear α decay case. This is the starting point in explaining a particular stock price behavior, as a tunnel effect.

Range bound markets very frequently encounter an "explosive" price movement out of the support or resistance levels. These spectacular moves of the stock price are now assimilated to a tunneling effect. What is the probability of the price to tunnel out of the range bound?

### 3. Stock price transmission coefficient

Asking the above question is like asking: what is the probability of a particle to penetrate out of the potential walls? The answer is to be found in a relation called price transmission coefficient.

The financial article I review here uses in deducing the price transmission coefficient the same road that is employed in nuclear physics for α particle decay. This approach may be questioned. It can be argued that is inappropriate to model the price behavior on the market. That will be argues based on no explicit content.

Leaving aside this debate for now, I will point out that the transmission coefficient for the stock price is found to be:

$$T = e^{-2\sqrt{\frac{r}{\sigma^4}(\sigma^2+r)}\left[\frac{1}{2}\ln\left(\left|\frac{\sqrt{1-\frac{r}{\sigma}K^2}+1}{\sqrt{1-\frac{r}{\sigma}K^2}-1}\right|\right)-\sqrt{1-\frac{r}{\sigma}K^2}\right]} \qquad (4)$$

where K is the strike price of the option, the limit beyond that the option is called to be in the money. The relation looks complicated but is extremely easy to compute for every market practitioner. It is no need to know quantum mechanics to find the probability of price tunneling if you simply know the relation (4).

The relation for T is kind of different for the α decay one not only because of the particularities of the applied domain but also because of the potential $V_{(S)}$ shape, still the meaning of T remains the same: probability of price tunneling out of the potential walls.

Empirical tests of price tunneling assumption led to the conclusion that T must be over 95% to see the tunneling effect happens. The T value is straightforward related to the width of the walls the stock price should penetrate through and also with the **λ** constant value. As relation (1) shows **λ** depend on interest rate and volatility. Since interest rate is relatively stable along of an option contract life, so tunneling of price is strongly dependent on the stock volatility.

Turning back to Figure 1it is easy to see that a higher **λ** means a lower wall width for the price to penetrate. Following the logic to the end it means that the stock volatility should fall dramatically prior of a stock tunneling effect in order to have a higher **λ**, and of course, a higher transmission coefficient value.

The dramatic fall in the stock volatility is a defining element for the tunnel effect in stock market; some example can be seen in Figure 2.

Figure 2.

Notice from the figure above, the abrupt fall in the volatility along with an "explosive" upside move of the stock price.

### 4. Philosophical debate

I already talk along this review article about philosophical aspects related to use of nuclear α decay theory to phenomena from financial markets. I also addressed some questions I encounter from individuals that came across my article "Time independent pricing of options in range bound markets", along the time. Many of these questions, the one in the title included, will probably remain unanswered for an extended period of time.

I should evidence here the fact that the time independent pricing of options equation was naturally deduced as a consequence of particular price moving in the stock market, namely a range bound market. It certainly is not an attempt to predict future market price, and also is not an attempt to model all aspects of market. It may not be relevant for the correctness of the option valuation by time independent equation, but certainly is in its favor.

From the various questions I came across with, I will retain one that is preponderant in discussions involving Schrodinger equation for price behavior: How could Schrodinger equation be involved in financial market since it is well known that it only models atomic level of Nature? Wrong! This is just a misunderstood perception of quantum world. Quantum models can be applied in all aspects of Nature, but their effects are observable only at the atomic level, having the small particles in the principal role. Except for financial markets, where similar effects can be seen with our own eyes, as it is reveled in the present reviewed article.

The importance of article findings may be at a small scale for particle physics, but what physicist can resist to the temptation of see with its very own eyes a tunneling effect in action, even this is happening in stock market and not in a Uranium nucleus?

I commence this review article with a question and I will end it with a question: why should Nature "invent" new rules for market price behavior if it already explain similar phenomenon by quantum theory?

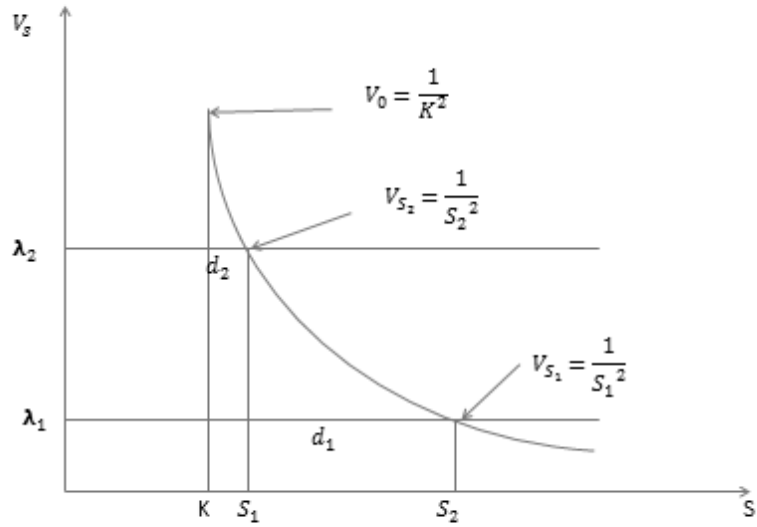

Figure 1. Shape of the stock price potential and different levels for **λ** constant.

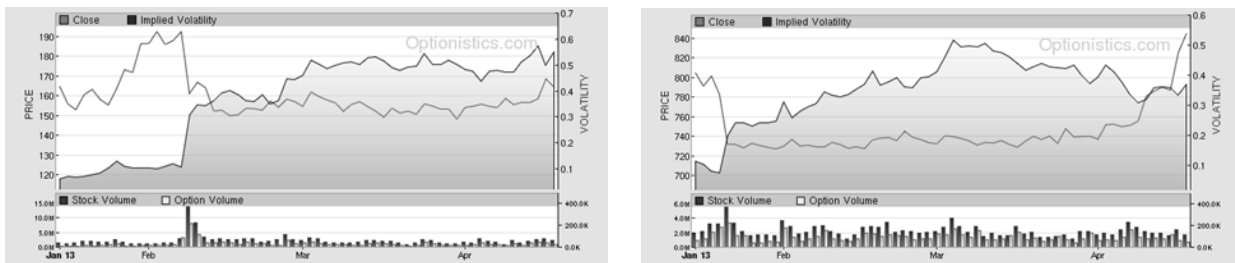

Figure 2. The fall in the stock volatility preceding the tunneling price effect for LNKD(up) and GOOG(down) stocks.

7